\documentclass[10pt,a4paper]{article}

\usepackage[normalem]{ulem}
\usepackage{cite}
\usepackage{color}
\usepackage{graphicx}
\usepackage[usenames,dvipsnames]{xcolor} 

\usepackage[]{array}
\usepackage[]{amsmath}
\usepackage[]{amssymb}
\usepackage[]{mathrsfs}
\usepackage[]{tensor}
\usepackage{bm}

\newcommand{\be}{\begin{eqnarray}}
\newcommand{\ee}{\end{eqnarray}}

\begin{document}

\begin{flushright}
{ZTF-EP-15-01}
\end{flushright}
\vskip 1cm

\begin{center}

{\LARGE{\bf Parity-odd surface anomalies\\[1mm] and \\[4mm] correlation functions on conical
defects}}
\vskip 2cm

{\Large M.~Cvitan$^a$, P.~Dominis Prester$^b$, S.~Pallua$^a$,
I.~Smoli\'c$^a$, T.~\v{S}temberga$^a$}\\
{}~\\
\quad \\
{\em ~$~^{a}$Theoretical Physics Division of Particles and Fields},\\
{\em Department of Physics, Faculty of Science, University of Zagreb},\\
{\em Bijeni\v{c}ka cesta 32, 10000 Zagreb, Croatia}
{}~\\
\quad \\
{\em ~$~^{b}$Department of Physics, University of Rijeka,}\\
{\em  Radmile Matej\v{c}i\'{c} 2, 51000 Rijeka, Croatia}\\
\vskip 1cm
Email: mcvitan@phy.hr, pprester@phy.uniri.hr, pallua@phy.hr, ismolic@phy.hr, tstember@phy.hr

\end{center}

\vskip 2cm 

\noindent
{\bf Abstract.}

\medskip

\noindent
We analyse the parity-odd (``type P'') surface anomalies of the energy-momentum tensor correlators in conformal field theories, with an emphasis on $d=4$ and $d=3$ dimensional spacetimes. Using cohomology analysis we construct the expression for the most general P-type surface trace anomaly on a singular 2-dimensional surface in 4-dimensional bulk spacetimes. As an important example, we specialise to the case when the singular surface is a conical defect and show that the bulk P-type Pontryagin trace anomaly induces such a surface trace anomaly. We show that this conical type P surface trace anomaly is given purely by the outer curvature tensor. In addition, we analyse parity-odd surface contact terms in energy-momentum tensor correlators in the flat spacetime induced by the conical defect by studying two special cases in which the contact terms are induced by, (1) type P trace anomaly in $d=4$ and, (2) gravitational Chern-Simons Lagrangian term in $d=3$ spacetime dimensions. In both cases we show that the surface contact terms appear in correlators of the lower rank than the corresponding bulk surface terms.

\vskip 1cm 

 
\vfill\eject

\section{Introduction}
\label{sec:intro}


It is known that when a quantum field theory is defined on the curved spacetime, an expectation value of the energy-momentum tensor may receive quantum corrections in the form of local terms. Some of these terms break classical symmetries of the theory, the well-known example being the trace (also called Weyl) anomaly in conformal field theories in even number $2k$ of the spacetime dimensions
\cite{Capper:1974ic,Deser:1976yx,Bonora:1985cq,Deser:1993yx,Nakayama:2012gu,Nakayama:2013is}:
\be \label{trangen}
\langle T_\mu^\mu(x) \rangle_g = \mathcal{A}(x)
\ee 
Here $T_{\mu\nu}$ is the energy-momentum tensor of the theory, and $\mathcal{A}$ is the trace anomaly which is generally some combination of monomials of $k$-th order in the Riemann tensor. The terms in the anomaly $\mathcal{A}$ are, therefore, local covariant expressions constructed from the background metric $g_{\mu\nu}$, the Riemann tensor and covariant derivatives.\footnote{Our conventions are as follows. The Riemann tensor is $R^\mu{}_{\alpha\nu\beta} = \partial_\nu \Gamma^\mu{}_{\alpha\beta} - \ldots$ and the Ricci tensor is $R_{\mu\nu} = R^\mu{}_{\alpha\mu\beta}$. The energy-momentum tensor is defined by 
$T_{\mu\nu} = 2 g^{-1/2} \delta S/\delta g^{\mu\nu}$, where $S$ is the action, and so $\langle T_{\mu\nu} \rangle = 2 g^{-1/2} \delta W/\delta g^{\mu\nu}$ where the functional $W[g]$ is defined by $W[g] = - \ln Z[g] = - \ln \int \mathcal{D} \phi\, \exp (-S)$. All calculations in this paper are performed in the Euclidean time.} $\mathcal{A}$ may come in three types: type A which is the $k$-th Euler invariant, type B which consists of contracted tensor products of the Weyl tensor, and type P consisting of exterior products of the Riemann curvature two-forms. The types A and B are parity-even, while type P is parity-odd and, although allowed by consistency conditions, was until recently usually neglected in the literature. 

Last decade witnessed a renewed interest in gravitational mechanisms of CP violation, see e.g.,
\cite{Alexander:2009tp,Bonora:2011gz,Bonora:2011mf,Bonora:2012xv,Bonora:2012eb,Bonora:2013jca,%
Alexander:2014bsa,Azeyanagi:2014sna,Mauro:2014eda}, including studying possibilities and consequences of the appearance of the type P anomalies \cite{Bonora:1985cq,Solodukhin:2005ns,Nakayama:2012gu,%
Closset:2012vp,Banerjee:2012cr,Loganayagam:2012zg,Bonora:2014qla,Bonora:2015nqa}. Indeed, in 
\cite{Bonora:2014qla,Bonora:2015nqa} an old result of \cite{Duff:78} was rederived and it was shown that in 4-dimensional quantum field theories with chiral fermions in which the numbers of left and right chiralities are not the same, type P anomalies are indeed present.

It is of interest to study non-regular spacetimes containing singular submanifolds. Singular structure may arise from the matter localised on the surface or from the topology, example of the former being branes and of the latter orbifolds. In some instances, possibly after analytical extensions, these singular surfaces are equivalent to the conical defects. One notable example is provided by the replica method \cite{Callan:1994py} for the calculation of the entanglement entropy (for reviews see \cite{Ryu:2006ef,Solodukhin:2011gn}), in which one effectively (by analytical continuation) introduces the conical singularity on the entangling surface (which is a codimension-2 submanifold) with a deficit parameter $\alpha$ and calculates the linear term in the expansion in $(1-\alpha)$.

In the presence of a singular surface $\Sigma$ the trace anomaly receives also surface contributions localised on $\Sigma$, an example of which is given by the so called Graham-Witten anomalies \cite{Graham:1999pm}. In $d=4$ dimensions a generic cohomology analysis of the parity-even surface trace anomalies connected with codimension-2 singular surfaces was performed in \cite{Schwimmer:2008yh}. One purpose of this paper is to extend this analysis to the parity-odd sector and in this way complete the construction from \cite{Schwimmer:2008yh}. We also calculate the parity-odd surface trace anomaly in the important case of conical singularity in $d=4$.

Singular surfaces generally also induce surface contact terms in correlation functions of the energy-momentum tensor in the flat spacetime. In fact, as was observed in the case of the conical singularity in \cite{Solodukhin:2013yha}, in the flat spacetime surface contact terms appear in the correlation functions one point lower than the corresponding bulk contact terms. This is an interesting property which may have some important uses in the future. We analyse and demonstrate this feature in the parity-odd sector on the two important examples containing a codimension-2 singular surface, (1) surface contact terms induced by the surface trace anomaly in $d=4$, and (2) surface contact terms induced by the presence of the gravitational Chern-Simons terms in the quantum effective action in $d=3$ when the singular surface is a conical singularity. We perform calculations using two methods, first developed in \cite{Solodukhin:2014dva} and the second in \cite{Smolkin:2014hba}. A side result of agreement of the two calculations is a non-trivial confirmation of the validity of both methods in the parity-odd sector.


\section{Bulk and surface trace anomalies}
\label{sec:surftran}


\subsection{Trace anomalies in general}
\label{ssec:tagen}


Let us assume that an otherwise regular (``bulk'') spacetime $M$ with the metric $g_{\mu\nu}$ contains a singular surface $\Sigma$, for which in this paper we shall assume to be codimension-2. A consequence is that some local quantities get contributions localised on $\Sigma$. Let us focus on a trace anomaly of a CFT defined on such a spacetime,\footnote{We assume Euclidean time, and so spacetimes will be Riemannian in this paper.}
\be
\mathcal{A}_\omega = 2\, \delta_\omega \ln Z[g] = \int_M \omega\, \langle T_\mu^\mu \rangle
\ee
where $Z[g]$ is the generating functional of the CFT on the spacetime with the metric $g_{\mu\nu}$, and $\omega(x)$ is an infinitesimal local parameter of the Weyl transformation $\delta g_{\mu\nu} = \omega\, g_{\mu\nu}$. In the presence of a singular surface, the trace anomaly in general receives the bulk and the surface contributions,
\be 
\mathcal{A}_\omega = \mathcal{A}_\omega^{(b)} + \mathcal{A}_\omega^{(\Sigma)} \quad ,
\ee
which are local functionals defined as
\be
\mathcal{A}_\omega^{(b)} &=& \int_M \omega\, \mathcal{A}^{(b)}
\nonumber \\
\mathcal{A}_\omega^{(\Sigma)} &=& \int_\Sigma \omega\, \mathcal{A}^{(\Sigma)} + \int_\Sigma \nabla^\mu \omega\, \mathcal{A}^{(\Sigma)}_\mu + \cdots
\ee
The dots $\cdots$ above denote the terms which include the second or the higher derivatives of $\omega$. Observe that in the case of the surface terms one cannot in general shift the derivatives acting on $\omega$ to the anomaly density by using partial integration, as is always possible for the bulk term.

In general, possible terms that can appear in the trace anomaly are restricted by the dimension, and by the consistency conditions. For the diffeomorphism covariant theories, in which the diff-anomaly is vanishing, consistency conditions reduce to
\be \label{trcr}
\delta_\omega \mathcal{A}_\omega = 0 \quad ,
\ee
where $\omega$ is treated as an anticommuting variable satisfying $\delta_\omega \omega = 0$. The ``true'' anomaly consists of the terms which are not exact with respect to $\delta_\omega$, i.e.~one has to subtract all the terms which can be written as $\delta_\omega \mathcal{C}$, where $\mathcal{C}$ is some diff-covariant density.

Cohomology analysis of the bulk anomalies has been thoroughly studied and it is known that possible terms in $\mathcal{A}^{(b)}$ fall into three-classes: type A, consisting of the Euler densities; type B, consisting of the Weyl-invariant terms; type P, consisting of the parity-odd terms. In $d=4$, the most general form for the bulk trace anomaly is:
\begin{equation} \label{tran4D}
\mathcal{A}^{(b)} = - \frac{a}{64}\, E_2 + \frac{c}{64}\, (W_{\mu\nu\rho\sigma})^2 + p\, P_2
\end{equation}
where $a$, $c$ and $p$ are constants depending on the CFT in question, $E_2$ is the second Euler scalar (Gauss-Bonnet scalar), $W_{\mu\nu\rho\sigma}$ is the Weyl tensor, and $P_2$ is the Pontryagin (pseudo)scalar
\begin{equation} \label{Pontr}
P_2 = \frac{1}{2\sqrt{g}}\, \varepsilon^{\mu\nu\rho\sigma}\,
R_{\alpha\beta\mu\nu}\, R^{\alpha\beta}{}_{\rho\sigma} \quad .
\end{equation}
Here, $\varepsilon^{\mu\nu\rho\sigma}$ is the Levi-Civita symbol (with components equal 0 or $\pm 1$). The first two terms in (\ref{tran4D}) are the type A and the type B bulk trace anomalies, respectively, which are parity-even, while the third term is the parity-odd type P bulk trace anomaly.

As for the surface anomalies, the cohomology analysis is not known in such detail as in the bulk case. The situation is more involved as it depends on the dimensionality of $\Sigma$ and also because building blocks include both extrinsic and intrinsic geometric quantities which complicates constructions. Again, surface anomalies can be divided in type A, B and P, in the same manner as for the bulk part.


\subsection{Codimension-2  surface trace anomalies in $d=4$}
\label{ssec:c2d4sa}


Let us now focus our attention on a particular case of the codimension-2 surface anomalies in $d=4$, where $\Sigma$ is a 2-dimensional singular surface. The relevant geometric objects include the first fundamental form (i.e.~the induced metric) $\gamma_{\mu\nu}$ and the corresponding intrinsic Ricci curvature scalar $\hat{R}$, the Weyl tensor contracted with the induced metric $W$, the second fundamental form (i.e.~the extrinsic curvature) $K_{\mu\nu\alpha}$ and the outer curvature pseudoscalar $\Omega$.\footnote{The mathematical definitions and properties of these objects can be found in \cite{Carter:2000wv,Cao:2010vj}. In our notation the extrinsic curvature is 
$K_{\mu\nu\rho}= \gamma_{\mu}{}^{\alpha}\gamma_{\nu}{}^{\beta}\nabla_\alpha \gamma_{\beta\rho}$. 
The outer curvature $ \Omega_{\mu\nu\rho\sigma}$ is defined using 
$(\tilde{D}_\mu \tilde{D}_\nu - \tilde{D}_\nu \tilde{D}_\mu)X_\rho = \Omega_{\mu\nu\rho\sigma}X^\sigma$, 
for $X^\mu$ normal i.e.~$\gamma^{\mu}{}_{\nu} X^\nu= 0$.
Here, the derivative $\tilde{D}$ is defined to act on normal vectors as 
$\tilde{D}_\mu X_\nu = \gamma_{\mu}{}^{\alpha} n_{\nu}{}^{\beta}\nabla_\alpha X_\beta$, and on tangential-normal $X_{\beta\sigma}$ (where $n_{\mu}{}^{\beta} X_{\beta\sigma} = 0$ and 
$\gamma_{\rho}{}^{\sigma} X_{\beta\sigma} = 0$) as  $\tilde{D}_\mu X_{\nu\rho} = \gamma_{\mu}{}^{\alpha}\gamma_{\nu}{}^{\beta} n_{\rho}{}^{\sigma}\nabla_\alpha X_{\beta\sigma}$.} From the dimensional analysis, it is easy to see that higher order curvatures are irrelevant here.

The cohomology of the parity-even sector was analysed and classified in \cite{Schwimmer:2008yh} with the result that the terms with derivatives of $\omega$ are not present and $\mathcal{A}^{(\Sigma)}$ is built out of the following densities:
\be \label{typeAsan}
&&\textrm{(type A) :} \qquad \hat{E}_1 = \hat{R}
 \\
&&\textrm{(type B) :} \qquad W = \gamma^{\mu\rho} \gamma^{\nu\sigma}\, W_{\mu\nu\rho\sigma} 
\quad , \quad (C_{\mu\nu\alpha})^2 = (K_{\mu\nu\alpha})^2 - \frac{1}{2} (K_\alpha)^2
\label{typeBsan}
\ee
So, the general form of the parity-even surface trace anomaly in this case is:
\begin{equation} \label{istape}
\big( \mathcal{A}_\omega^{(\Sigma)} \big)_{\textrm{even}}
 = \int_\Sigma \omega \big [k_1 \hat{E}_1 + k_2 W + k_3 (C_{\mu\nu\alpha})^2 \big]
\end{equation}
where $k_a$ are the coefficients that depend both on the theory and on the properties of the singular surface.

We want to complete the analysis by finding the most general expression for the parity-odd (type P) surface trace anomaly. Now, it is not hard to see that there are just three linearly independent candidate terms which have the proper dimension:
\be \label{d4sacan}
\int_\Sigma \omega\, \Omega \ \ ,\quad \int_\Sigma \omega\, \epsilon^{\mu\nu\alpha\beta} K_{\mu\rho\alpha} K_{\nu}{}^{\rho}{}_{\beta} \qquad \textrm{and}\qquad \int_\Sigma \epsilon^{\alpha\beta}\, \nabla_\alpha \omega\, K_{\beta} \ \ .
\ee
Here $\epsilon_{\alpha\beta}$ is the binormal on $\Sigma$ and $\Omega$ is the outer curvature pseudoscalar, defined on $\Sigma$, which is obtained from the outer curvature tensor $\Omega_{\mu\nu\alpha\beta}$ through\footnote{Note that $\Omega_{\mu\nu\alpha\beta}$ has only one independent component, so $\Omega$ contains the complete information about the outer curvature tensor.}
\begin{equation} \label{omegats}
\Omega \equiv \frac{1}{2} \epsilon^{\mu\nu\alpha\beta}\, \Omega_{\mu\nu\alpha\beta}
\end{equation}

The first and the second term in (\ref{d4sacan}) indeed satisfy (\ref{trcr}), but the third term does not and it drops out. It may appear that there is one more potential candidate
\be \label{cand}
\epsilon^{\mu\nu\rho\sigma} \gamma_\mu^\alpha\, \gamma_\nu^\beta\, 
n_\rho^\lambda\, n_\sigma^\kappa\, R_{\alpha\beta\lambda\kappa}
\ee
where $n_{\mu\nu} = g_{\mu\nu} - \gamma_{\mu\nu}$, which indeed satisfies the consistency condition, but it is not independent and so we leave it out. One can see this by using the Ricci equation
\begin{equation} \label{ricci}
\Omega_{\mu\nu\rho\sigma} = \gamma_\mu^\alpha\, \gamma_\nu^\beta\,
n_\rho^\lambda\, n_\sigma^\kappa\, R_{\alpha\beta\lambda\kappa}
+ K_{\mu\tau\rho} K_\nu{}^\tau{}_\sigma - K_{\mu\tau\sigma} K_\nu{}^\tau{}_\rho
\end{equation}
which contracted with the Levi-Civita tensor gives
\begin{equation} \label{POm}
\epsilon^{\mu\nu\rho\sigma} \gamma_\mu^\alpha\, \gamma_\nu^\beta\, 
n_\rho^\lambda\, n_\sigma^\kappa\, R_{\alpha\beta\lambda\kappa}  
= 2 \, \Omega - 2\, \epsilon^{\mu\nu\rho\sigma} K_{\mu\tau\rho} K_{\nu}{}^{\tau}{}_{\sigma}
\end{equation}
We see that the candidate (\ref{cand}) can be written as a linear combination of those in (\ref{d4sacan}).

So, our final result of this section is that in the parity-odd sector there are no terms containing derivatives of $\omega$, and 
$\mathcal{A}^{(\Sigma)}$ is built out of the following two terms:
\be 
\textrm{(type P) :} \qquad \Omega \qquad,\qquad
 \epsilon^{\mu\nu\alpha\beta} K_{\mu\rho\alpha} K_{\nu}{}^{\rho}{}_{\beta} \qquad
\ee
which means that the \emph{general form of the parity-odd surface trace anomaly} in $d=4$ on 2-dimensional singular surfaces is given by
\be \label{d4icsa}
\big( \mathcal{A}_\omega^{(\Sigma)} \big)_{\textrm{odd}} = \int_\Sigma \omega
 \left( \tilde{k}_1 \Omega + \tilde{k}_2\, \epsilon^{\mu\nu\alpha\beta} K_{\mu\rho\alpha} K_{\nu}{}^{\rho}{}_{\beta} \right) \ .
\ee
Again, the coefficients $\tilde{k}_{1,2}$ in general depend both on the theory in hand and on the properties of the singular surface $\Sigma$. In the next section we shall calculate them for the particular case when $\Sigma$ is a conical defect surface.


\section{Surface trace anomalies for conically singular surfaces}
\label{sec:condef}


\subsection{Conical defects}
\label{sec:condef}


We want to present an explicit example where the type P surface trace anomalies are present, and for this we take an important example of spacetimes with conical singularities.

We assume that in otherwise regular $d$-dimensional spacetime $\mathcal{M}_0$ with metric $g_{\mu\nu}$ a conical defect with an angle deficit $2\pi (1-\alpha)$ is introduced in a standard fashion such that there is a $(d-2)$-dimensional singular surface $\Sigma$ (the ``tip of the cone''). The spacetime with a conical singularity is denoted by $\mathcal{M}_\alpha$, and the singular surface $\Sigma$ by 
$\mathcal{C}$. When necessary, we shall use the local coordinates $x^\mu$, $\mu=1,\ldots,d$ in which $\Sigma$ is defined with $x_1=x_2=0$ and the conical defect is described by having an angle deficit in $x_1$-$x_2$ plane.

We shall be interested in the integrals over $\mathcal{M}_\alpha$ of local functions $F$ of the curvature scalars constructed out of $g^{\mu\nu}$, Levi-Civita tensor $\epsilon_{\mu_1\cdots\mu_d}$, Riemann tensor $R_{\mu\nu\rho\sigma}$ and covariant derivatives 
$\nabla_\mu$. It was shown in \cite{Dong:2013qoa} that in the leading order in $2\pi (1-\alpha)$ such integrals can be split into the bulk part (integral over $\mathcal{M}_0$) and the surface part (integral over $\Sigma$) by using the following formula
\begin{eqnarray} \label{dongf}
\int_{\mathcal{M}_\alpha}\!\! F \!&=&\! \int_{\mathcal{M}_0}\!\! F \, + 2\pi (1 - \alpha)
 \int_\mathcal{C} \left\{ \frac{\partial F}{\partial R_{\mu\rho\nu\sigma}}\,
 \epsilon_{\mu\rho}\, \epsilon_{\nu\sigma} - 2 \sum_r \left(
 \frac{\partial^2 F}{\partial R_{\mu_1\rho_1\nu_1\sigma_1}\, \partial R_{\mu_2\rho_2\nu_2\sigma_2}} \right)_{\!r} 
  \frac{K_{\rho_1 \sigma_1 \lambda_1}\, K_{\rho_2 \sigma_2 \lambda_2}}{q_r + 1} \right.
\nonumber \\
   &&  \times \left[(n_{\mu_1 \mu_2}\, n_{\nu_1 \nu_2} - \epsilon_{\mu_1 \mu_2}\, \epsilon_{\nu_1 \nu_2})\, n^{\lambda_1 \lambda_2} 
   + (n_{\mu_1 \mu_2}\, \epsilon_{\nu_1 \nu_2} + \epsilon_{\mu_1 \mu_2}\, n_{\nu_1 \nu_2})\, \epsilon^{\lambda_1 \lambda_2} \right]
    \bigg\} + O\big((1-\alpha)^2\big)
\end{eqnarray}
Summation over $r$ and the definition of the parameter $q_r$ are explained in \cite{Dong:2013qoa}.  As in our examples $F$ will be linear or quadratic in Riemann tensor, we can put $q_r=0$. We shall be interested only in the lowest-order correction in the expansion over $(1-\alpha)$.

It is useful to introduce two orthonormal vector fields, $n^\mu_{(a)}$, $a=1,2$, which constitute a basis in the subspace of vectors normal to $\Sigma$. Then one can write
\begin{equation}
n^{\mu\nu} = \sum_{a=1}^2 n^\mu_{(a)} n^\nu_{(a)} \qquad, \qquad
\epsilon_{\mu\nu} = \sum_{a,b=1}^2 n^\mu_{(a)} n^\nu_{(b)}\, \varepsilon_{ab}
\end{equation}
where $\varepsilon_{ab}$ is the two-dimensional Levi-Civita symbol. Using this it is easy to prove a useful relation
\begin{equation}
\epsilon_{\mu\nu} \epsilon_{\rho\sigma} = n_{\mu\rho} n_{\nu\sigma} - n_{\mu\sigma} n_{\nu\rho}
\end{equation}


\subsection{Conical trace anomaly in $d=4$}
\label{sec:4Dcorr}


We now focus on the CFT defined on the four dimensional curved spacetime $\mathcal{M}_\alpha$ which contains the conical singularity located on the 2 dimensional surface $\Sigma = \mathcal{C}$, as described in the previous section. As the trace anomaly (\ref{tran4D}) has a purely geometric description in a regular spacetime, we assume that it preserves its form also in the presence of the conical singularity, in the sense of the regularisation of the spacetime already implicit in Dong's formula 
(\ref{dongf}). This means that the integrated anomaly is given by:
\begin{equation} \label{d4trafull}
\mathcal{A}_\omega = \int_{\mathcal{M}_\alpha} \omega
 \left( - \frac{a}{64}\, E_2 + \frac{c}{64}\, (W_{\mu\nu\rho\sigma})^2 + p\, P_2 \right)
\end{equation}

We want to extract the surface contributions to the trace anomaly. In the cases of the type A and B conical surface trace anomalies the results are known and can be found in \cite{Solodukhin:2008dh,Fursaev:2013fta}. Here we shall complete the analysis in $d=4$ by calculating the type P conical surface trace anomalies. The easiest way to achieve this is by applying Dong's formula (\ref{dongf}) on (\ref{d4trafull}). A tedious but straightforward calculation gives the following result
\begin{equation} \label{anomPc}
\big(\mathcal{A}_\omega^{(\mathcal{C})}\big)_\textrm{odd} = 4\pi\, p\,(1-\alpha)
 \int_\mathcal{C} \omega\, \epsilon^{\mu\nu\rho\sigma} \left( \gamma_\mu^\alpha\,
 \gamma_\nu^\beta\,  n_\rho^\lambda\, n_\sigma^\kappa\, R_{\alpha\beta\lambda\kappa}
 + 2\, K_{\mu\tau\rho} K_{\nu}{}^{\tau}{}_{\sigma} \right)
\end{equation}
The first and the second term separately come from the first and the second term in (\ref{dongf}), respectively. By using (\ref{POm}) we can write our result for the integrated type P conical surface anomaly in the more compact and suggestive form using just the outer curvature tensor
\begin{equation} \label{anomPcf}
\big(\mathcal{A}_\omega^{(\mathcal{C})}\big)_\textrm{odd} = 8\pi\, p\,(1-\alpha)
 \int_\mathcal{C} \omega\, \Omega 
\end{equation}
The pseudotensor $\Omega$ was defined in (\ref{omegats}). Comparison with the generic formula (\ref{d4icsa}) for the surface trace anomaly in $d=4$ shows that for the conical anomaly one has $\tilde{k}_1 = 8\pi\, p\,(1-\alpha)$ and $\tilde{k}_2 = 0$.\footnote{Between the first and the second version of our paper a reference \cite{Azeyanagi:2015uoa} appeared, with some results partially overlapping with ours. Though the motivation and notation are different, the results there are in agreement with ours.} We see that in the case of the conical singularity, the parity-odd surface trace anomaly is a direct consequence of a presence of the parity-odd bulk trace anomaly, and is completely determined by it. As already mentioned, there is now a strong evidence \cite{Bonora:2014qla,Bonora:2015nqa} that CFT's with chiral fermions indeed possess such a bulk anomaly. 

It is interesting to note that the outer curvature scalar $\Omega$ can be written as the total 2-dimensional gradient, and so is in some sense an outer analogue of the Euler term in $d=2$ (which is an intrinsic Ricci scalar) \cite{Carter:2000wv}. As the conical surface type A anomaly is purely given by the intrinsic Ricci scalar, it is maybe not surprising that the type P anomaly is given purely by $\Omega$. Moreover, Penrose showed that generally the spinor approach leads naturally to the construction of a single complex curvature invariant on 2-dimensional submanifolds whose real part is the Euler term (the intrinsic Ricci scalar) and imaginary part is the outer curvature scalar $\Omega$ \cite{Penrose:1987uia}. We believe that there is some interesting mathematics here worthy of detail studying, but we leave such questions to our future research.

\section{Correlation functions in a flat spacetime}
\label{sec:cfflat}

As is well-known, the local terms in the 1-point energy-momentum correlation function on a curved spacetime induce contact terms in higher-rank energy-momentum correlation functions on a flat space. Our definition of correlation functions in the flat space is such that
\begin{equation} \label{emcfdef}
\langle T_{\mu_1\nu_1}(x_1) \ldots T_{\mu_n \nu_n}(x_n)\, T_{\mu\nu}(x) \rangle = (-2)^n
 \frac{\delta}{\delta g^{\mu_1 \nu_1}(x_1)} \cdots \frac{\delta}{\delta g^{\mu_n \nu_n}(x_n)}\,
  \langle T_{\mu\nu}(x) \rangle_g \bigg|_{g_{\mu\nu} = \delta_{\mu\nu}}
\end{equation}
Of course, if $\langle T_{\mu\nu}(x) \rangle$ contains both bulk and surface local terms there will be corresponding bulk and surface contact terms. Eq. (\ref{emcfdef}) is telling us that in the case of the surface contact terms generated by the surface trace anomaly in $d=4$ one should expect from (\ref{istape}) and (\ref{d4icsa}) that they will appear already in the 2pt energy-momentum correlation functions, while from (\ref{tran4D}) one expects that the bulk contact terms start at the 3pt functions. 
The fact that in general number of dimensions d, the surface contact terms start to appear in the correlation functions of one rank less than the bulk terms was noted in the reference 
\cite{Solodukhin:2013yha} and then elaborated in \cite{Solodukhin:2014dva} in the case of the contact terms connected with the parity even conical surface trace anomalies. This also follows from Dong’s formula (\ref{dongf}), which we will use for studying parity odd surface contact terms.

In this section we further demonstrate this phenomenon, of surface contact terms appearing ``before" bulk surface terms in energy-momentum tensor correlation functions, on two examples. One example is the contact term in $d=4$ connected with the type P surface trace anomaly, and the other is the specific contact term in $d=3$ appearing when an effective action contains the gravitational Chern-Simons term. Let us emphasize that the correlators defined in (\ref{emcfdef}) are, strictly speaking, not equal to standard $T$-ordered correlation functions. However, the difference is irrelevant for the calculation of the contact terms which we present in the rest of the section.

\subsection{Conical surface contact terms in $d=4$}
\label{ssec:4Dcorrf}


\subsubsection{Method 1}
\label{sssec:4Dm1}


According to (\ref{emcfdef}), the presence of the bulk trace anomalies in $d=4$, given in (\ref{tran4D}), obviously induces bulk contact terms in the energy-momentum tensor correlation functions of the type
\begin{equation} \label{d4gcf}
\langle T_{\mu_1\nu_1}(x_1)\cdots T_{\mu_n \nu_n}(x_n)\, T_\rho^\rho(x) \rangle = (-2)^n\,
 \frac{\delta}{\delta g^{\mu_1 \nu_1}(x_1)}\cdots \frac{\delta}{\delta g^{\mu_n \nu_n}(x_n)}\,
  \langle T_\rho^\rho(x) \rangle_g \bigg|_{g_{\mu\nu} = \delta_{\mu\nu}}
\end{equation}
when $n\ge 2$. In addition, when the singular surface $\Sigma$ is present one expects also that the surface contact terms are present. In the parity-odd sector we showed that in this case the trace anomaly is given by
\begin{equation}
\left( \langle T_\rho^\rho(x) \rangle_g \right)_{\mathrm{odd}} = p\, P_2(x) +  \left( \tilde{k}_1\, \Omega(x)
 + \tilde{k}_2\, \epsilon^{\mu\nu\alpha\beta} K_{\mu\rho\alpha} K_{\nu}{}^{\rho}{}_{\beta} \right) \delta_\Sigma
\end{equation}
Here $\delta_\Sigma$ is the covariant Dirac $\delta$-function localised on the 2-surface $\Sigma$. 

We shall now show that the term proportional to $\tilde{k}_1$ induces nonvanishing surface contact terms in (\ref{d4gcf}) also for $n=1$ correlation functions. To calculate this contribution we use the relation (\ref{POm}) and note that the term proportional to 
$\tilde{k}_2$ obviously does not contribute in (\ref{d4gcf}) with $n=1$, because $K_{\mu\rho\alpha} = 0$ in the flat spacetime. Using the formula
\begin{equation} \label{varRiem}
\frac{\delta R_{\alpha\beta\gamma\kappa}(x')}{\delta g^{\mu\nu}(x)}
\bigg|_{g_{\mu\nu}=\delta_{\mu\nu}}
 = - \frac{1}{4} (\delta_{\alpha\mu}\, \delta_{\kappa\nu}\, \partial_{\gamma}
\partial_{\beta}
  + \delta_{\beta\mu}\, \delta_{\gamma\nu}\, \partial_{\kappa}
\partial_{\alpha}) \delta^{(4)}(x-x')
  + (\mu \leftrightarrow \nu) - (\gamma \leftrightarrow \kappa)
\end{equation}
and the fact that in the flat space $R_{\mu\nu\rho\sigma} = 0$, we obtain
\begin{eqnarray} \label{TTdefres}
\left( \langle T_{\mu\nu}(x)\, T_\rho^\rho(x') \rangle \right)_{\mathrm{odd}} = - \tilde{k}_1\, \varepsilon_{\mu\gamma\rho\sigma}\, 
n_\nu^\rho\, n_\beta^\sigma\, \delta_\Sigma\, \partial^\beta \partial^\gamma \delta^{(4)}(x-x') + (\mu \leftrightarrow \nu)
\end{eqnarray}
Now we choose the coordinates such that $\Sigma$ is defined by $x_1 = x_2 = 0$. Then, $\delta_\Sigma = \delta(x_1) \delta(x_2)$ and we can take $n^\mu_{(1)} = (1,0,0,0)$ and $n^\mu_{(2)} = (0,1,0,0)$. Using all this
(\ref{TTdefres}) becomes
\begin{equation} \label{TTdefres2}
\left( \langle T_{\mu\nu}(x)\, T_\rho^\rho(x') \rangle \right)_{\mathrm{odd}}
 = -\tilde{k}_1\, \tilde{\varepsilon}_{\mu a}\, \hat{\varepsilon}_{\nu
\hat{a}}\, \delta(x_1) \delta(x_2)\, \partial^a \partial^{\hat{a}}
\delta^{(4)}(x-x') + (\mu \leftrightarrow \nu) + O((1-\alpha)^2)
\end{equation}
where $\hat{a}=1,2$ denotes directions normal to $\Sigma$, while $a=3,4$ denotes directions tangential to $\Sigma$, and 
$\hat{\varepsilon}$ and $\tilde{\varepsilon}$ are 2-dimensional Levi-Civita symbols living on normal and tangential space, respectively (the binormal and the volume form on $\Sigma$, respectively)
\begin{equation} \label{htveps}
\hat{\varepsilon}_{12} = 1 \;,\qquad \hat{\varepsilon}_{\mu a} = 0 \;,\qquad
 \tilde{\varepsilon}_{34} = 1 \;,\qquad \tilde{\varepsilon}_{\mu \hat{a}} = 0
\end{equation}

In the special case of the conical surface trace anomaly we have shown that
\begin{equation} \label{tk1con}
\tilde{k}_1 = 8\pi(1-\alpha)\, p
\end{equation}

From the expression (\ref{TTdefres2}) it is obvious that the correlation function is nonvanishing only if one of the indices $\mu$ or $\nu$ is normal while the other one is tangential to $\Sigma$. One of the consequences is that the trace of  (\ref{TTdefres2}) vanishes
\begin{equation}
\left( \langle T_\mu^\mu(x)\, T_\nu^\nu(x') \rangle \right)_{\mathrm{odd}} = 0
\end{equation}
As it was shown in \cite{Solodukhin:2014dva} that the same is true for the type B anomaly, the only contribution to the trace of the $n=1$ correlation function comes from the type A anomaly.


\subsubsection{Method 2}
\label{sssec:4Dm2}


As argued in \cite{Smolkin:2014hba}, there is a correspondence between the correlation functions on the flat space with and without the conical defect. This can be used to check the result from the previous section for the case of the conical surface trace anomaly. The correspondence is given through the relation
\begin{equation} \label{wwocone}
\mathcal{P} \langle \mathcal{O}_1(x_1) \ldots \mathcal{O}_N(x_N)
\rangle_{\mathcal{M}_\alpha}
 = \langle \mathcal{O}_1(x_1) \ldots \mathcal{O}_N(x_N) K_0 \rangle
\end{equation}
where
\begin{equation} \label{Poper}
\mathcal{P} = - \lim_{\alpha\to1} \frac{\partial}{\partial\alpha}
\end{equation}
and the operator $K_0$ is
\begin{equation} \label{K0def}
K_0 = -2\pi \int d^{d-2}y \int_0^\infty dz_1 z_1 T_{22}(z_1,z_2=0,\vec{y})
\end{equation}
Here, the directions 1 and 2 are normal while the directions $3,\ldots,d$ are tangential to the conical defect surface $\Sigma$, and 
$\vec{y}=(z_3,\ldots,z_d)$.

Applying this to the particular case of the parity-odd contribution to the correlation functions of the energy-momentum tensor in $d=4$ enables us to check the result from the last subsection by an independent method. From (\ref{wwocone}) follows
\begin{equation} \label{wwoc4d}
\mathcal{P} \left( \langle T_\rho^\rho(x)\, T_{\mu\nu}(y) \rangle_{\mathcal{M}_\alpha}
 \right)_{\mathrm{odd}} = -2\pi \int dz_3 \int dz_4 \int_{0}^{\infty} dz_1 z_1 \langle T_\rho^\rho(x)\,
T_{\mu\nu}(y) T_{22}(z_1,z_2=0,z_3,z_4) \rangle
\end{equation}
For clarity, we denote the correlation functions on the spacetime with the conical singularity with the subscript $\mathcal{C}_\alpha$. Now, from (\ref{tran4D}) it can be shown that in the regular flat spacetime one has 
\begin{eqnarray} \label{TTTpodd}
\langle T_\rho^\rho(x)\, T_{\mu\nu}(y)\, T_{\rho\sigma}(z) \rangle_{\mathrm{odd}}
 &=& 2p\, \varepsilon_{\mu\rho\alpha\beta}
 (\partial_\sigma \partial_\nu - \delta_{\sigma\nu} \partial^2 )
[\partial^\alpha \delta^{(4)}(x-y)\, \partial^\beta \delta^{(4)}(x-z)] 
\nonumber \\
 && +\, (\rho\leftrightarrow\sigma) + (\mu\leftrightarrow\nu) 
\end{eqnarray}
where the differential operator inside the round brackets is explicitly given by
\begin{equation} \label{deroper}
\partial_\sigma \partial_\nu - \delta_{\sigma\nu} \partial^2
 \equiv \frac{\partial}{\partial y_\sigma}
\frac{\partial}{\partial z_\nu}
  - \delta_{\sigma\nu} \frac{\partial}{\partial y_\kappa}
\frac{\partial}{\partial z_\kappa}
\end{equation}
Plugging (\ref{TTTpodd}) into (\ref{wwoc4d}) we obtain
\begin{eqnarray}
\mathcal{P} \left( \langle T_\rho^\rho(x)\, T_{\mu\nu}(y) \rangle_{\mathcal{M}_\alpha}
\right)_{\mathrm{P-odd}}
 \!\!&=&\!\! -8\pi\, p\, \varepsilon_{\mu2\alpha\beta} \int_{-\infty}^{\infty} dz_3
\int_{-\infty}^{\infty} dz_4  
 \int_{0}^{\infty} dz_1 z_1 \times 
\nonumber \\
&& \times(\partial_2\partial_\nu - \delta_{2\nu}\,
\partial_\sigma\partial_\sigma)
 \left[ \partial_\alpha \delta^{(4)}(x-y)\, \partial_\beta \delta^{(4)}(x-z)
\right] + (\mu \leftrightarrow \nu)
\label{wwoc4dv2}
\end{eqnarray}
Let us concentrate on the first two lines in (\ref{wwoc4dv2}). Observe that $\beta\ne 3$ or 4 because of the integrations over $z_3$ and $z_4$, and $\beta\ne2$ due to the Levi-Civita symbol, so it must be that $\beta=1$. From this follows that $\mu$ and $\alpha$ must be tangential to the surface of the defect, i.e., $\mu=3$ and $\beta=4$ or vice versa. Moreover, it can be shown that $\nu\ne 3$ or 4 because of the integration over $z_1$. Taking all this into account, and using
\begin{equation} \label{intz1}
\int_{0}^{\infty} dz_1 z_1 \frac{\partial^2}{\partial z_1^2} \delta(x_1 - z_1) =
\delta(x_1)
\end{equation}
it is easy to show that (\ref{wwoc4dv2}) becomes
\begin{eqnarray}
\mathcal{P} \left( \langle T_\rho^\rho(x)\, T_{\mu\nu}(y) \rangle_{\mathcal{M}_\alpha}
\right)_{\mathrm{odd}}
= -8\pi\, p\, \tilde{\varepsilon}_{\mu a}\, \hat{\varepsilon}_{\nu \hat{a}}\,
 \delta(x_1) \delta(x_2)\, \partial^{\hat{a}} \partial^a \delta^{(4)}(x-y) +
(\mu \leftrightarrow \nu)
\end{eqnarray}
where as before $\hat{a}=1,2$, $a=3,4$, and the 2-dimensional Levi-Civita symbols are those defined in (\ref{htveps}).
From (\ref{wwoc4dv2}) and (\ref{Poper}) and the fact that the bulk part of the 2-point correlation function vanishes in the limit $\alpha\to1$ it follows that integration over $\alpha$ gives in the lowest order in $(1-\alpha)$ the following result
\begin{equation} \label{TTdefres3}
\left( \langle T_\rho^\rho(x)\, T_{\mu\nu}(y) \rangle_{\mathcal{M}_\alpha}
\right)_{\mathrm{odd}}
= -8\pi(1-\alpha)\, p\, \tilde{\varepsilon}_{\mu a}\, \hat{\varepsilon}_{\nu
\hat{a}}\,
 \delta(x_1) \delta(x_2)\, \partial^{\hat{a}} \partial^a \delta^{(4)}(x-y) +
(\mu \leftrightarrow \nu)
\end{equation}
where we have used 2-dimensional Levi-Civita symbols defined in (\ref{htveps}). As expected, the result is the same as the corresponding one obtained by the Method 1 in the Sec. \ref{sssec:4Dm1} (see Eq. (\ref{TTdefres2})-(\ref{tk1con})).


\subsection{Conical surface contact terms in $d=3$}
\label{ssec:3Dcorr}


\subsubsection{Method 1}
\label{sssec:3Dm1}


When 3-dimensional QFT's are defined in a curved spacetime, expectation value of the energy-momentum tensor may develop a parity-odd contribution of the form
\begin{equation} \label{d3T}
\langle T_{\mu\nu}(x) \rangle_{\mathrm{odd}} = \frac{i w}{48\pi}\,
\varepsilon_{\alpha\beta(\mu} \nabla^\beta R^\alpha{}_{\nu)}
 = - \frac{i w}{48\pi}\,C_{\mu\nu}
\end{equation}
where $C_{\mu\nu}$ is known as the Cotton-York tensor. As the integer part of the coefficient $w$ can be removed by adding to the classical action a local counterterm, which is the well-known gravitational Chern-Simons term, it is sometimes stated that $w$ is defined modulo 1 \cite{Closset:2012vp}.\footnote{Arguments based on the path integral quantisation suggest that (i) the coupling constant of the purely gravitational Chern-Simons Lagrangian term in all odd spacetime dimensions is imaginary in the Euclidean regime, and (ii) the value of the coupling is quantised \cite{Bonora:2012xv}. The parametrisation used in (\ref{d3T}) is such that the contribution to $w$ from the  Lagrangian gravitational CS term must be an integer \cite{Witten:2007kt} if the only restriction on the spacetime is that it is a spin manifold.} It is known that in regular spacetimes the Cotton-York tensor is traceless and covariantly conserved
\begin{equation}
C_\mu{}^\mu = 0 \;,\qquad \nabla_\mu C^{\mu\nu} = 0
\end{equation}
so as a consequence (\ref{d3T}) does not contribute to the trace anomaly, which is expected from the general theorem stating that there are no trace anomalies in CFT's defined in odd-dimensional spacetimes.

Now we add into the spacetime a conical defect with the deficit angle $2\pi(1-\alpha)$ in the same manner as before. By assuming that (\ref{d3T}) is valid also when the conical defect is present, and using
\begin{equation}
(R_{\mu\nu})_{\mathcal{C}_\alpha} = R_{\mu\nu} + 2\pi (1-\alpha)\, n_{\mu\nu}\, \delta_\Sigma
 + \textrm{(terms containing the second fundamental form)}
\end{equation}
which follows from Dong's formula (\ref{dongf}), we obtain that for the flat metric 
$g_{\mu\nu} = \delta_{\mu\nu}$
\begin{eqnarray}
\left( \langle T_{\mu\nu}(x) \rangle_{\mathcal{M}_\alpha} \right)_{\mathrm{odd}}
 &=& \frac{i w}{24} (1-\alpha)\, \varepsilon_{\alpha\beta(\mu}\, \partial^\beta \!\left( n^\alpha_{\nu)}\, \delta_\Sigma \right)
\nonumber \\
 &=& \frac{i w}{48} (1-\alpha)\, \hat{\varepsilon}_{\nu\hat{a}} \delta_{\mu 3} \partial^{\hat{a}} [\delta(x_1) \delta(x_2)]
  + (\mu \leftrightarrow \nu)
\label{d3Talphaf}
\end{eqnarray}
where in the second equality we used (\ref{htveps}) and also the vanishing of the second fundamental form. Again, this expression is nonvanishing only if one of the indices is in the normal direction (1 or 2) while the other one is in the tangential direction (3). It is easy to see that (\ref{d3Talphaf}) is traceless and covariantly conserved. This means that, as expected, it will not contribute to the trace anomaly.


\subsubsection{Method 2}
\label{sssec:3Dm2}


For a CFT defined in a regular flat spacetime a consequence of (\ref{d3T}) is
\cite{Nakayama:2012gu,Closset:2012vp}
\begin{equation} \label{d3TT}
\langle T_{\mu\nu}(x)\,T_{\alpha\beta}(y) \rangle_{\mathrm{odd}} = \frac{i w}{192\pi}\,
\epsilon_{\mu\alpha\sigma}
 ( \partial_\nu \partial_\beta - \delta_{\nu\beta} \partial^2) \partial^\sigma
\delta^{(3)}(x-y)
 + (\mu \leftrightarrow \nu) + (\alpha \leftrightarrow \beta)
\end{equation}
Now we can use the correspondence (\ref{wwocone}) to independently calculate the expectation value of the energy momentum tensor in a flat space with a conical singularity. In this way we obtain
\begin{equation} \label{Talpha}
\mathcal{P} \langle T_{\mu\nu}(x) \rangle_{\mathcal{M}_\alpha} =  \langle
T_{\mu\nu}(x) K_0 \rangle
 = - 2\pi \int_{-\infty}^\infty dy_3 \int_0^\infty dy_1\, y_1\, \langle
T_{\mu\nu}(x)\,T_{22}(y1,y_2=0,y_3) \rangle
\end{equation}
By using (\ref{d3TT}) we obtain
\begin{eqnarray}
\langle T_{\mu\nu}(x) K_0 \rangle_{\mathrm{odd}}
  &=& - \frac{i w}{48}\, \varepsilon_{\mu2\sigma} (\partial_\nu \partial_2 -
\delta_{\nu2} \partial^2)  \int_{-\infty}^\infty dy_3
  \int_0^\infty dy_1\, y_1\left. \partial^\sigma \delta^{(3)}(x-y)
\right|_{y_2=0}
  \nonumber \\
 && +\, (\mu \leftrightarrow \nu)
\end{eqnarray}
Now, $\sigma\ne 3$ and $\nu\ne 3$ because integration over $y_3$ would be vanishing. From this follows that $\sigma=1$ and one gets
\begin{eqnarray*}
\langle T_{\mu\nu}(x) K_0 \rangle_{\mathrm{odd}}
  &=& \frac{i w}{48}\, \delta_{\mu 3} (\partial_\nu \partial_2 - \delta_{\nu2}
\partial^a \partial_a) \left[ \delta(x_2)
  \int_0^\infty dy_1\, y_1\, \partial^1 \delta(x_1-y_1) \right] + (\mu
\leftrightarrow \nu)
\nonumber \\
 &=& \frac{i w}{48}\, \delta_{\mu 3} \hat{\varepsilon}_{\nu \hat{a}} \partial^{\hat{a}}
[\delta(x_1) \delta(x_2)] + (\mu \leftrightarrow \nu)
\nonumber \\
 &=& \frac{i w}{48}\, \varepsilon_{\mu\alpha\beta}\, n^\alpha_\nu\, \partial^\beta
\delta_\Sigma + (\mu \leftrightarrow \nu)
\end{eqnarray*}
where $a=1,2$ and $\delta_\Sigma = \delta(x_1) \delta(x_2)$ in the particular Cartesian coordinates we use here. Plugging this into (\ref{Talpha}), using (\ref{Poper}) and integrating over $\alpha$ we obtain that in the leading order in $(1-\alpha)$ the final result is
\begin{eqnarray}
\left( \langle T_{\mu\nu}(x) \rangle_{\mathcal{M}_\alpha} \right)_{\mathrm{odd}}
 &=& \frac{i w}{48} (1-\alpha)\, \delta_{\mu 3} \hat{\epsilon}_{\nu a} \partial^a
[\delta(x_1) \delta(x_2)] + (\mu \leftrightarrow \nu)
\nonumber \\
 &=& \frac{i w}{48} (1-\alpha)\, \epsilon_{\alpha\beta\mu}\, n^\alpha_\nu\,
\partial^\beta \delta_\Sigma
 + (\mu \leftrightarrow \nu)
\nonumber \\
 &=& \frac{i w}{24} (1-\alpha)\, \epsilon_{\alpha\beta(\mu}\, \partial^\beta \!\left( n^\alpha_{\nu)}\, \delta_\Sigma \right)
\label{Talphaf}
\end{eqnarray}
We see that the final result is the same as the one obtained by the Method 1, which is (\ref{d3Talphaf}).


\section{Conclusion}
\label{sec:concl}

The expectation value of the energy-momentum tensor on a curved spacetime may receive quantum corrections in the form of local terms. The well-know examples are the trace and diff- anomalies, which break the classical symmetries of the theory. In some instances, e.g.~in theories in four dimensions with unequal number of left and right chiral fermions, the parity breaking gravitational terms appear.

We have studied some consequences of the presence of singular surfaces for quantum field theories in curved spacetimes, focusing on the parity violating sector. In particular, we have studied a parity-odd contribution to the trace anomaly of the conformal field theories, which we dubbed the type P surface trace anomaly. By analysing the consistency condition we have found the most general form for the type P surface trace anomaly in four dimensions. To have an example at hand, we have analysed the special case when the singular surface is due to conical singularities, which are important in its own right, and obtained the exact result for the type P conical surface trace anomaly. This surface anomaly appears only when the theory contains parity violating bulk trace anomaly and it turns out that it can be expressed fully by the outer curvature tensor, an interesting property which deserves further studying.

In the second part of the paper we have studied, on the two examples, the influence of singular surfaces on the parity violating contact terms in energy-momentum correlation functions on the flat spacetime. One example consists of the surface contact terms in four dimensions connected with the type P surface trace anomalies. The second example consists of the surface contact terms generated by the presence of the gravitational Chern-Simons term in the effective action in the three dimensional flat spacetime. In the case of the conical singularity, we were able to perform calculations by using two methods, one purely geometrical and the other using the connections between correlation functions in the flat spacetime with and without conical singularity. The agreement of the results gives a non-trivial confirmation of the validity of both methods in the parity-odd sector.

Generalisations of our analyses to higher spacetime dimensions are possible, as type P trace anomalies are expected to be present in $4k$ dimensions, while gravitational Chern-Simons terms in effective actions are expected to be present in $(4k-1)$ dimensions 
\cite{Solodukhin:2005ns,Bonora:2011mf,Bonora:2013jca}. However, computations become much more complicated so we left this to the future work.

\vspace{15pt}

\noindent
{\bf \large Acknowledgements}

\bigskip

\noindent 
The research has been supported by Croatian Science Foundation under the project No.~8946 and by University of Rijeka under the research support No.~13.12.1.4.05. We thank Loriano Bonora on many lessons on anomalies and Bruno Lima de Souza for stimulating discussions.

\vspace*{5pt}






\begin{thebibliography}{99}


\bibitem{Capper:1974ic}
  D.~M.~Capper and M.~J.~Duff,
  ``Trace anomalies in dimensional regularization,''
  Nuovo Cim.\ A {\bf 23} (1974) 173.

\bibitem{Deser:1976yx}
  S.~Deser, M.~J.~Duff and C.~J.~Isham,
  ``Nonlocal Conformal Anomalies,''
  Nucl.\ Phys.\ B {\bf 111} (1976) 45.

\bibitem{Bonora:1985cq}
  L.~Bonora, P.~Pasti and M.~Bregola,
  ``Weyl Cocycles,''
  Class.\ Quant.\ Grav.\  {\bf 3} (1986) 635.

\bibitem{Deser:1993yx}
  S.~Deser and A.~Schwimmer,
  ``Geometric classification of conformal anomalies in arbitrary dimensions,''
  Phys.\ Lett.\ B {\bf 309} (1993) 279
  [hep-th/9302047].

\bibitem{Nakayama:2012gu}
  Y.~Nakayama,
  ``CP-violating CFT and trace anomaly,''
  Nucl.\ Phys.\ B {\bf 859} (2012) 288
  [arXiv:1201.3428 [hep-th]].

\bibitem{Nakayama:2013is}
  Y.~Nakayama,
  ``Scale invariance vs conformal invariance,''
  Phys.\ Rept.\  {\bf 569} (2015) 1
  [arXiv:1302.0884 [hep-th]].

\bibitem{Alexander:2009tp}
  S.~Alexander and N.~Yunes,
  ``Chern-Simons Modified General Relativity,''
  Phys.\ Rept.\  {\bf 480} (2009) 1
  [arXiv:0907.2562 [hep-th]].

\bibitem{Bonora:2011gz}
  L.~Bonora, M.~Cvitan, P.~Dominis Prester, S.~Pallua and I.~Smoli\'{c},
  ``Gravitational Chern-Simons Lagrangians and black hole entropy,''
  JHEP {\bf 1107} (2011) 085
  [arXiv:1104.2523 [hep-th]].

\bibitem{Bonora:2011mf}
  L.~Bonora, M.~Cvitan, P.~D.~Prester, S.~Pallua and I.~Smoli\'{c},
  ``Gravitational Chern-Simons Lagrangian terms and spherically symmetric spacetimes,''
  Class.\ Quant.\ Grav.\  {\bf 28} (2011) 195009
  [arXiv:1105.4792 [hep-th]].

\bibitem{Bonora:2012xv}
  L.~Bonora, M.~Cvitan, P.~D.~Prester, S.~Pallua and I.~Smoli\'{c},
  ``Gravitational Chern-Simons terms and black hole entropy. Global aspects,''
  JHEP {\bf 1210} (2012) 077
  [arXiv:1207.6969 [hep-th]].

\bibitem{Bonora:2012eb}
  L.~Bonora, M.~Cvitan, P.~D.~Prester, S.~Pallua and I.~Smoli\'{c},
  ``Stationary rotating black holes in theories with gravitational Chern-Simons Lagrangian term,''
  Phys.\ Rev.\ D {\bf 87} (2013) 024047
  [arXiv:1210.4035 [hep-th]].

\bibitem{Bonora:2013jca}
  L.~Bonora, M.~Cvitan, P.~Dominis Prester, S.~Pallua and I.~Smoli\'{c},
  ``Symmetries and gravitational Chern-Simons Lagrangian terms,''
  Phys.\ Lett.\ B {\bf 725} (2013) 468
  [arXiv:1305.0432 [hep-th]].

\bibitem{Alexander:2014bsa}
  S.~Alexander, S.~Cormack, A.~Marciano and N.~Yunes,
  ``Gravitational-Wave Mediated Preheating,''
  Phys.\ Lett.\ B {\bf 743} (2015) 82
  [arXiv:1405.4288 [gr-qc]].

\bibitem{Azeyanagi:2014sna}
  T.~Azeyanagi, R.~Loganayagam, G.~S.~Ng and M.~J.~Rodriguez,
  ``Covariant Noether Charge for Higher Dimensional Chern-Simons Terms,''
  JHEP {\bf 1505} (2015) 041
  [arXiv:1407.6364 [hep-th]].

\bibitem{Mauro:2014eda}
  S.~Mauro and I.~L.~Shapiro,
  ``Anomaly-induced effective action and Chern-Simons modification of general relativity,''
  Phys.\ Lett.\ B {\bf 746} (2015) 372
  [arXiv:1412.5002 [gr-qc]].

\bibitem{Solodukhin:2005ns}
  S.~N.~Solodukhin,
  ``Holographic description of gravitational anomalies,''
  JHEP {\bf 0607} (2006) 003
  [hep-th/0512216].

\bibitem{Closset:2012vp}
  C.~Closset, T.~T.~Dumitrescu, G.~Festuccia, Z.~Komargodski and N.~Seiberg,
  ``Comments on Chern-Simons Contact Terms in Three Dimensions,''
  JHEP {\bf 1209} (2012) 091
  [arXiv:1206.5218 [hep-th]].

\bibitem{Banerjee:2012cr}
  N.~Banerjee, S.~Dutta, S.~Jain, R.~Loganayagam and T.~Sharma,
  ``Constraints on Anomalous Fluid in Arbitrary Dimensions,''
  JHEP {\bf 1303} (2013) 048
  [arXiv:1206.6499 [hep-th]].

\bibitem{Loganayagam:2012zg}
  R.~Loganayagam,
  ``Anomalies and the Helicity of the Thermal State,''
  JHEP {\bf 1311} (2013) 205
  [arXiv:1211.3850 [hep-th]].

\bibitem{Bonora:2014qla}
  L.~Bonora, S.~Giaccari and B.~Lima de Souza,
  ``Trace anomalies in chiral theories revisited,''
	JHEP {\bf 1407}, 117 (2014)
  [arXiv:1403.2606 [hep-th]].

\bibitem{Bonora:2015nqa}
  L.~Bonora, A.~D.~Pereira and B.~Lima de Souza,
  ``Regularization of energy-momentum tensor correlators and parity-odd terms,''
  JHEP {\bf 1506} (2015) 024
  [arXiv:1503.03326 [hep-th]].

\bibitem{Duff:78} 
  S. M. Christensen and M. J. Duff,
  ''Axial and conformal anomalies for arbitrary spin in gravity and supergravity'', 
  Phys.\ Lett.\ {\bf 76B} (1978) 571. 

\bibitem{Callan:1994py}
  C.~G.~Callan, Jr. and F.~Wilczek,
  ``On geometric entropy,''
  Phys.\ Lett.\ B {\bf 333} (1994) 55
  [hep-th/9401072].

\bibitem{Ryu:2006ef}
  S.~Ryu and T.~Takayanagi,
  ``Aspects of Holographic Entanglement Entropy,''
  JHEP {\bf 0608} (2006) 045
  [hep-th/0605073].

\bibitem{Solodukhin:2011gn}
  S.~N.~Solodukhin,
  ``Entanglement entropy of black holes,''
  Living Rev.\ Rel.\  {\bf 14} (2011) 8
  [arXiv:1104.3712 [hep-th]].

\bibitem{Graham:1999pm}
  C.~R.~Graham and E.~Witten,
  ``Conformal anomaly of submanifold observables in AdS / CFT correspondence,''
  Nucl.\ Phys.\ B {\bf 546} (1999) 52
  [hep-th/9901021].

\bibitem{Schwimmer:2008yh}
  A.~Schwimmer and S.~Theisen,
  ``Entanglement Entropy, Trace Anomalies and Holography,''
  Nucl.\ Phys.\ B {\bf 801} (2008) 1
  [arXiv:0802.1017 [hep-th]].

\bibitem{Solodukhin:2013yha}
  S.~N.~Solodukhin,
  ``The a-theorem and entanglement entropy,''
  arXiv:1304.4411 [hep-th].

\bibitem{Solodukhin:2014dva}
  S.~N.~Solodukhin,
  ``Conformal a-charge, correlation functions and conical defects,''
  Phys.\ Lett.\ B {\bf 736} (2014) 283
  [arXiv:1406.5368 [hep-th]].

\bibitem{Smolkin:2014hba}
  M.~Smolkin and S.~N.~Solodukhin,
  ``Correlation functions on conical defects,''
  Phys.\ Rev.\ D {\bf 91} (2015) 4,  044008
  [arXiv:1406.2512 [hep-th]].

\bibitem{Carter:2000wv}
  B.~Carter,
  ``Essentials of classical brane dynamics,''
  Int.\ J.\ Theor.\ Phys.\  {\bf 40} (2001) 2099
  [gr-qc/0012036].

\bibitem{Cao:2010vj}
  L.~M.~Cao,
  ``Deformation of Codimension-2 Surface and Horizon Thermodynamics,''
  JHEP {\bf 1103} (2011) 112
  doi:10.1007/JHEP03(2011)112
  [arXiv:1009.4540 [gr-qc]].

\bibitem{Dong:2013qoa}
  X.~Dong,
  ``Holographic Entanglement Entropy for General Higher Derivative Gravity,''
  JHEP {\bf 1401} (2014) 044
  [arXiv:1310.5713 [hep-th], arXiv:1310.5713].


\bibitem{Fursaev:2013fta}
  D.~V.~Fursaev, A.~Patrushev and S.~N.~Solodukhin,
  ``Distributional Geometry of Squashed Cones,''
  Phys.\ Rev.\ D {\bf 88} (2013) 4,  044054
  [arXiv:1306.4000 [hep-th]].

\bibitem{Solodukhin:2008dh}
  S.~N.~Solodukhin,
  ``Entanglement entropy, conformal invariance and extrinsic geometry,''
  Phys.\ Lett.\ B {\bf 665} (2008) 305
  [arXiv:0802.3117 [hep-th]].

\bibitem{Azeyanagi:2015uoa}
  T.~Azeyanagi, R.~Loganayagam and G.~S.~Ng,
  ``Holographic Entanglement for Chern-Simons Terms,''
  arXiv:1507.02298 [hep-th].

\bibitem{Penrose:1987uia}
  R.~Penrose and W.~Rindler,
  ``Spinors and Space-time: Volume 1, Two-Spinor Calculus and Relativistic Fields,''
  Cambridge, Uk: Univ. Pr. (1984) (Cambridge Monographs On Mathematical Physics)

\bibitem{Witten:2007kt}
  E.~Witten,
  ``Three-Dimensional Gravity Revisited,''
  arXiv:0706.3359 [hep-th].


\end{thebibliography}
\end{document}